\def\beq{\begin{equation}}
\def\eeq{\end{equation}}
\documentclass{Rinton-P9x6}

\begin{document}

\title{Large Neutrino Mixing in Grand Unified Theories}

\author{F. Feruglio}

\address{University of Padova and I.N.F.N., Padova, Italy\\E-mail: feruglio@pd.infn.it}


\maketitle

\abstracts
{A non-minimal, semi-realistic version of supersymmetric ${\rm SU(5)}$
grand unified theory is discussed. The solution of the doublet-triplet splitting problem
leads to a better agreement between the predicted and observed values of the 
low-energy strong coupling constant and to a prolongation of the proton lifetime. A $U(1)$
flavor symmetry allows to accommodate a realistic mass spectrum in the charged and in the
neutral fermion sectors and protects doublet-triplet splitting and proton 
decay from dangerous radiative corrections or non-renormalizable operators.}

%\section{Guidelines}%1

Minimal versions of grand unified theories (GUTs) are plagued by severe fine-tuning problems.
\begin{itemize} 
\item{\sf Doublet-triplet splitting} -- By far the most severe problem, it requires
an unnatural adjustment of the superpotential parameters of one part in $10^{14}$. Moreover,
the tree-level solution to this problem can be spoiled either by radiative corrections when supersymmetry
(SUSY) is broken, or by non renormalizable operators.
\item{\sf Proton decay} -- Minimal ${\rm SU(5)}$ is ruled out by the recent SuperKamiokande data  
$\tau_p/BR(p\to K^+ {\bar\nu}) > 2\cdot 10^{33}$ ys (90\% CL) \cite{SK}. Moreover generic, non-renormalizable operators contributing to proton decay and originating at the Planck 
scale $M_{Pl}$ are expected to predict a proton lifetime of order $10^{20}$ ys.
\item{\sf Wrong mass relations} -- The equality of down quark and charged lepton masses at the GUT scale 
in minimal ${\rm SU(5)}$ is inaccurate for the first two generations, even though it is correct order-of-magnitude
wise. The whole neutrino sector is missing in minimal ${\rm SU(5)}$.
\item{\sf Strong coupling constant} -- Gauge coupling unification in minimal GUT 
leads to a value of $\alpha_s(M_Z)$ that, although affected by large uncertainties, tends to be too 
large: $\alpha_s(M_Z)=0.13\pm 0.01$ for colour triplets at the GUT scale and
supersymmetric particles close to 1 TeV. 
\end{itemize}
There can also be additional issues more specific to the supersymmetric realization
of GUT ideas, like for instance the supersymmetric flavour problem. In ref. \cite{afm2} we discuss 
a semi-realistic model that addresses and solves the above problems in an extended version
of ${\rm SU(5)}$, supplemented by a ${\rm U(1)_Q}$ flavour symmetry.

In this model the doublet triplet-splitting problem is solved by a variant of the missing partner mechanism
employing, beyond 5 and ${\bar 5}$, the 50, $\overline{50}$, $75\equiv Y$ and $1\equiv X$ ${\rm SU(5)}$
representations (see table 1) \cite{ber}. The multiplet $Y$, singlet under the flavour symmetry, breaks ${\rm SU(5)}$ down to ${\rm SU(3)}\times {\rm SU(2)}\times
{\rm U(1)}$, while $X$, characterized by $Q=-1$, is the only field charged under ${\rm U(1)_Q}$ that acquires a large VEV. In the limit of exact SUSY, the doublets in 5 and ${\bar 5}$ remain
massless. The mass $m_T$ of the effective triplet suppressing the dimension 5, $\vert\Delta B\vert=1$ operators
is proportional to $\langle Y\rangle^2/\langle X \rangle$, where $\langle X \rangle$ is undetermined.
Finally, operators like $5{\bar 5} X^mY^n$ $(m,n>0)$, which potentially could destabilize the doublets,
are forbidden by ${\rm U(1)_Q}$. When SUSY is broken, $\langle X \rangle$ acquires a VEV close to the cut-off
$\Lambda$ of the theory and a $\mu$ term can be generated \`a la Giudice-Masiero from a higher-dimensional
term in the K\"ahler potential.

The spectrum of heavy particles associated to the missing partner mechanism produces
two main effects.
\begin{itemize}
\item{} The strong coupling constant $\alpha_s(M_Z)$ receives large threshold corrections from the 
splitted $Y$ supermultiplet.
As a result, $\alpha_s(M_Z)$ is smaller than in minimal ${\rm SU(5)}$. Indeed
values of $m_T$ larger by a factor 20-30 than in minimal ${\rm SU(5)}$ are required to reconcile the 
prediction of $\alpha_s(M_Z)$ with the data, with a direct advantage for proton decay.
\item{} The model is no longer asymptotically free, due to the large field content. The ${\rm SU(5)}$
coupling constant blows up at a scale $\overline{\Lambda}$ smaller than the Planck scale. We typically
find $\Lambda\le\overline{\Lambda}~\approx 20~M_{GUT}$.
\end{itemize}
 
Fermion masses are obtained from the ${\rm U(1)_Q}$ charge assignment given in table 1.
\begin{table}[t]
\caption{Chiral Multiplets Quantum Numbers.}\label{tab:qn}
\begin{center}
\begin{tabular}{|c|c|c|} 
 
\hline 
 
\raisebox{0pt}[13pt][7pt]{Field} & 
 
\raisebox{0pt}[13pt][7pt]{${\rm SU(5)}$} & 

\raisebox{0pt}[13pt][7pt]{${\rm U(1)_Q}$}\\

\hline\hline

\raisebox{0pt}[13pt][7pt]{$H$} & 
 
\raisebox{0pt}[13pt][7pt]{5} & 
 
\raisebox{0pt}[13pt][7pt]{-2}\\ 
 
\hline

\raisebox{0pt}[13pt][7pt]{${\overline{H}}$} & 
 
\raisebox{0pt}[13pt][7pt]{${\overline{5}}$} & 
 
\raisebox{0pt}[13pt][7pt]{+1}\\ 
 
\hline

\raisebox{0pt}[13pt][7pt]{$H_{50}$} & 
 
\raisebox{0pt}[13pt][7pt]{50} & 
 
\raisebox{0pt}[13pt][7pt]{2}\\ 
 
\hline

\raisebox{0pt}[13pt][7pt]{$H_{\overline{50}}$} & 
 
\raisebox{0pt}[13pt][7pt]{${\overline{50}}$} & 
 
\raisebox{0pt}[13pt][7pt]{-1}\\ 
 
\hline

\raisebox{0pt}[13pt][7pt]{$Y$} & 
 
\raisebox{0pt}[13pt][7pt]{75} & 
 
\raisebox{0pt}[13pt][7pt]{0}\\ 
 
\hline

\raisebox{0pt}[13pt][7pt]{$X$} & 
 
\raisebox{0pt}[13pt][7pt]{1} & 
 
\raisebox{0pt}[13pt][7pt]{-1}\\ 
 
\hline\hline

\raisebox{0pt}[13pt][7pt]{$\Psi_{10}$} & 
 
\raisebox{0pt}[13pt][7pt]{10} & 
 
\raisebox{0pt}[13pt][7pt]{(4,3,1)}\\ 
 
\hline

\raisebox{0pt}[13pt][7pt]{$\Psi_{\bar 5}$} & 
 
\raisebox{0pt}[13pt][7pt]{${\bar 5}$} & 
 
\raisebox{0pt}[13pt][7pt]{(4,2,2)}\\ 
 
\hline

\raisebox{0pt}[13pt][7pt]{$\Psi_1$} & 
 
\raisebox{0pt}[13pt][7pt]{1} & 
 
\raisebox{0pt}[13pt][7pt]{(1,-1,0)}\\ 

\hline
\end{tabular}
\end{center}
\end{table}
As well known, abelian charges constrain the spectrum up to unknown coefficients 
of order one. It is possible to choose these coefficients in order to correctly reproduce
quark masses, mixing angles and the CP violating phase. The model predicts $\tan\beta
\approx O(1)$, which also moderates the proton decay amplitudes.
The neutrino sector of the model is quite similar to the one discussed in ref. \cite{af}.
As a consequence of the $U(1)$ assignment and of the see-saw mechanism, a large mixing
for atmospheric neutrinos is obtained. Such a mixing is directly related to 
a large mixing between the right-handed $s$ and $b$ quark fields, via
the minimal ${\rm SU(5)}$ relation $m_e={m_d}^T$, which is approximately valid also 
in the present model. This is the reason why a large mixing among leptons is
compatible with small quark mixing angles, even in a GUT, where lepton and quarks
belong to the same representations of the gauge group. 
The solar mixing angle is expected to be close to maximal and, numerically,
the so called LOW and vacuum oscillation solutions are equally possible.
Finally a $\theta_{13}$ angle of order $0.05$ is predicted.

A well known obstacle in minimal ${\rm SU(5)}$ is the strict equality $m_e={m_d}^T$,
compatible with the third generation, but inexact for the first and the second
families. The correction requires order-one adjustments
that can be obtained in the present model by allowing, beyond the minimal
$\Psi_{10} G_d \Psi_{\bar 5} {\bar 5}$ Yukawa coupling also the non-renormalizable
term $1/\Lambda \Psi_{10} F_d \Psi_{\bar 5} {\bar 5} Y$. The $Y$ multiplet
differentiates charged leptons from down quarks and we find:
\beq
m_d\approx \left[G_d+\frac{\langle Y\rangle}{\Lambda} F_d\right]~~~,
\eeq
\beq
m_e^T\approx \left[G_d-3\frac{\langle Y\rangle}{\Lambda} F_d\right]~,
\eeq
where the $3\times 3$ matrices $G_d$ and $F_d$ are constrained by the flavour symmetry.
It is interesting to observe that we can reproduce the relations $m_\tau\approx m_b$,
$m_\mu\approx 3 m_s$ and $m_e\approx  m_d/3$, by taking $\langle Y\rangle/\Lambda$ of 
order 0.1, in agreement with $\Lambda\le\overline{\Lambda}~\approx 20~M_{GUT}$.
While the predictivity of the model is reduced because non-renormalizable operators
are only suppressed by powers of $M_{GUT}/\Lambda$, still these corrections could explain
the small distortion of the spectrum with respect to the minimal model. 
  
Proton decay dominant amplitudes are derived from the dimension 5 superpotential:
\beq
w=\frac{1}{m_T}\left[ Q \hat{A} Q Q \hat{C} L 
+ U^c \hat{B} E^c U^c \hat{D} D^c\right]~,
\eeq
which, although formally equal to that of the minimal model, exhibits four important differences:
\begin{itemize}
\item{} An effective triplet mass $m_T$, larger by a factor 20-30 than in  minimal ${\rm SU(5)}$
leads to a suppression factor 400-900 in rate.
\item{} An additional Yukawa coupling
is allowed by the symmetries of the theory: $\Psi_{10} G_{\overline{50}} \Psi_{10} {\overline{50}}$. 
While the couplings of the conventional term
$\Psi_{10} G_u \Psi_{10} 5$ are restricted by the up quark masses, the couplings of
the new term are unconstrained, since $\langle{\overline{50}}\rangle=0$. We obtain:
\beq
\hat{B}=-2\hat{A}=\left[G_u-\frac{c_2 \langle Y\rangle}{c_4 \langle X\rangle} G_{\overline{50}}\right]~,
\eeq
where $c_2$ and $c_4$ are dimensionless coefficients. As a consequence, a large region in parameter space 
exists where a sizeable destructive interference between the $G_u$ and the $G_{\overline{50}}$ contributions 
can occur.
\item{}Also the $\hat{C}$ and $\hat{D}$ couplings are distorted:
\beq
\hat{C}=\left[-G_d-\frac{\langle Y\rangle}{\Lambda} F_d\right]~,
\eeq
\beq
\hat{D}=\left[G_d-\frac{\langle Y\rangle}{\Lambda} F_d\right]~.
\eeq
This modification, however, has a not-too-large effect on proton decay rates.
\item{}
The non-renormalizable operators that could originate at the cut-off scale $\Lambda$
are controlled by the flavour symmetry and lead to a contribution to the proton decay
amplitude that can be comparable to the one coming from the color triplet exchange.
\end{itemize}
As a result, the predicted range for the proton decay rates considerably extends
with respects to that of minimal ${\rm SU(5)}$, allowing values that are not incompatible
with the present limits and are testable in the next generation of experiments.
In particular, our numerical estimate gives
$8\cdot 10^{31}~{\rm ys}<\tau_p/BR(p\to K^+ {\bar\nu})<3\cdot 10^{34}~{\rm ys}$ 
and $2\cdot 10^{32}~{\rm ys}<\tau_p/BR(p\to \pi^+ {\bar\nu})<8\cdot 10^{34}~{\rm ys}$. 

In summary, it is a remarkable feature of the model that the presence of the 
representations 50, ${\overline{50}}$ and 75, demanded by the missing 
partner mechanism for the solution of the doublet-triplet splitting
problem, directly produces, through threshold corrections at $M_{GUT}$,
a decrease of the value of $\alpha_s(m_Z)$ that corresponds to
coupling unification and an increase in the effective mass that mediates
proton decay. As a consequence the value of the strong coupling is in
better agreement with the experimental value and the proton decay
rate is smaller by a factor 400-900 than in the minimal model.
The presence of these large representations also has the consequence
that the asymptotic freedom of ${\rm SU(5)}$ is spoiled
and the associated gauge coupling becomes non perturbative below
$M_{Pl}$. We argue that this property far from being unacceptable can 
actually play an important role to obtain better results for
fermion masses.

\section*{Acknowledgments}
I warmly thank Guido Altarelli and Isabella Masina for the enjoyable collaboration on which this talk is based. The present work has been partially 
supported by the European Program HPRN-CT-2000-00148 
(network Across The Energy Frontier).

\end{document}